# Real-time imputation of missing predictor values in clinical practice

Nijman SWJ[a], Hoogland J[a], Groenhof TKJ[a], Brandjes M[b], Jacobs JJL[b], Bots ML[a], Asselbergs FW[cde], Moons KGM[a], Debray TPA[ae]

On behalf of the UCC-CVRM and UCC-SMART study groups.

a Julius Center for Health Sciences and Primary Care, University Medical Center Utrecht, Utrecht University, Utrecht, The Netherlands;

b Department of Health, Ortec B.V. Zoetermeer, The Netherlands;

c Department of Cardiology, University Medical Center Utrecht, Utrecht University, The Netherlands;

d Institute of Cardiovascular Science, Faculty of Population Health Sciences, University College London, London, United Kingdom;

e Health Data Research UK, Institute of Health Informatics, University College London, London, United Kingdom

Corresponding author: Steven WJ Nijman, Julius Center for Health Sciences and Primary Care, University Medical Center Utrecht, Heidelberglaan 100, 3584 CX, Utrecht, the Netherlands.

**Abstract**

**Introduction –** Use of prediction models is widely recommended by clinical guidelines, but usually requires complete information on all predictors that is not always available in daily practice.

**Methods –** We describe two methods for real-time handling of missing predictor values when using prediction models in practice. We compare the widely used method of mean imputation (M-imp) to a method that personalizes the imputations by taking advantage of the observed patient characteristics. These characteristics may include both prediction model variables and other characteristics (auxiliary variables). The method was implemented using imputation from a joint multivariate normal model of the patient characteristics (joint modeling imputation; JMI). Data from two different cardiovascular cohorts with cardiovascular predictors and outcome were used to evaluate the real-time imputation methods. We quantified the prediction model's overall performance (mean squared error (MSE) of linear predictor), discrimination (c-index), calibration (intercept and slope) and net benefit (decision curve analysis).

**Results –** When compared with mean imputation, JMI substantially improved the MSE (0.10 vs. 0.13), c-index (0.70 vs 0.68) and calibration (calibration-in-the-large: 0.04 vs. 0.06; calibration slope: 1.01 vs. 0.92), especially when incorporating auxiliary variables. When the imputation method was based on an external cohort, calibration deteriorated, but discrimination remained similar.

**Conclusions –** We recommend JMI with auxiliary variables for real-time imputation of missing values, and to update imputation models when implementing them in new settings or (sub)populations.

# 1 Introduction

The identification and treatment of patients at increased risk for disease is a cornerstone of personalized and stratified medicine (1–3). Often, identification of high-risk patients involves the use of multivariable risk prediction models. These models combine patient and disease characteristics to provide estimates of absolute risk of a disease in an individual (4–8). For example, prediction models for cardiovascular disease such as Framingham heart score (FHS) (9), HEART score (1), ADVANCE (10), Elderly (11) and SMART (12) are well known examples (13). Additionally, cardiovascular guidelines recommend use of prediction models integrated in computerized decision support systems (CDSS), to support guideline adherent, risk-informed decision making (1,13).

When applying a risk prediction model in real-time, which constitutes its application in individual patients in routine clinical practice, one needs to have the individual's information (values) on all predictors in the model. Otherwise no absolute risk prediction by the model can be generated, restricting its use in situations when a physician is unable to acquire certain patient measurements. For example, for cardiovascular risk assessment, prediction models require complete information typically on age, sex, smoking, co-morbidities, blood pressure and lipid levels (14). With the increased availability of large databases with information from electronic health care records, automated implementation and use of risk prediction models within CDSS using routine care (EHR) data has gained much interest (15–19). However, the use of EHR databases faces many challenges, notably the incompleteness of data in the records (19–22). The usability of a prediction model may thus still be limited in clinical practice if its implementation cannot standardly handle missing predictor values in real time. A detailed example is given in Box 1.

A variety of strategies have been developed for daily practice to handle missing predictor values in real-time(23,24). Imputation strategies are of interest since they allow for direct use of well-known prediction models in their original form. In short, imputation substitutes a missing predictor value with one or more plausible values (imputations). In its simplest form, these imputations solely rely on the estimated averages of the missing variables in the targeted population. Therefore, they reflect what is known about the average patient. These simple methods can be applied directly in real-time clinical practice, provided that summary information (e.g. mean predictor values) about the targeted population is directly available. Additionally, imputations can account for the individual patient's *observed* predictor values by making use of the estimated associations between the patient characteristics in other patients. In that case, the imputations reflect all what is known about the specific individual at hand. Usually, the implementation of more complex imputation strategies requires direct access to the raw data from multiple individuals, which is typically problematic in clinical practice (e.g. due to operational or privacy constraints). As such, alternative strategies are required to make the imputation model applicable in real-time clinical practice.

Although real-time imputation of missing predictor values in clinical practice offers an elegant solution to generate predictions in the presence of incomplete data, the accuracy of these predictions may be severely limited if imputed values are a poor representation of the unobserved predictor values. In particular, problems may arise when (i) the imputation procedure does not adequately leverage information from the observed patient data, and (ii) if the estimated population characteristics used to generate the imputation(s) poorly represent the population to which the individual patient belongs. It is currently unclear how these novel real-time imputation methods influence the accuracy of available prediction models.

In this paper we explicitly focus on the relatively new area of real-time imputation, which has not been studied often before in similar literature. Most similar studies that address missing data consider and attempt to halt the onset of missing data in a particular dataset with missing values in study individuals, rather than a missing predictor in a single individual that is encountered in real-time clinical practice. Briefly, we investigate the performance of these two real-time imputation methods to handle missing predictor values when using a prediction model in daily practice. We evaluated both the accuracy of imputation and the impact of imputation on the prediction model's performance. Furthermore, transportability of the imputation procedures across different populations was empirically examined in two cardiovascular cohorts.

## 2 Methods

### 2.1 Short description

We conducted a simulation study to evaluate the impact of real-time imputation of missing predictor values on the absolute risk predictions in routine care. Hereto, we considered 2 large datasets and two real-time imputation methods. The datasets considered were the ongoing Utrecht Cardiovascular cohort - Cardiovascular risk management (UCC-CVRM) and the Utrecht Cardiovascular cohort - Secondary Manifestation of ARTerial disease (UCC-SMART) study (25,26). Both studies focused on cardiovascular disease prevention and included newly referred patients visiting the University Medical Center (UMC) Utrecht for evaluation of cardiovascular disease (25,26). Baseline examinations (i.e. predictors) for the UCC-CVRM included only the minimum set as suggested by the Dutch Cardiovascular Risk Management Guidelines (27).

### 2.2 Imputation methods

We considered mean imputation (M-Imp) and joint modelling imputation (JMI) (28,29). Mean imputation was chosen as a comparison due to its straightforward implementation and extensive use during prediction model development and validation (30–33). A major advantage of mean imputation is that it does not require information on individual patient characteristics and can be implemented without much difficulty in daily clinical practice. Using mean imputation, missing predictor values are simply imputed by their respective mean, usually from a representative sample (e.g. observational study). JMI was chosen because it allows to personalize imputations by adjusting for observed characteristics. To this purpose, JMI implements multivariate methods that have extensively been studied in the literature (28,29,34,35). Some modifications are required to implement JMI for real-time imputation, these have been discussed previously (23). In JMI, missing predictor values are imputed by taking the expected value from a multivariate distribution that is conditioned on the observed patient data. Implementations of JMI commonly assume that all variables are normally distributed, as this greatly simplifies the necessary calculations. This method then minimally requires mean and covariance estimates for all variables that are included as predictors in the prediction model from a representative sample (e.g. observational study). As an extension to JMI, we also consider that additional patient data (auxiliary variables) are available and can be used to inform the imputation of missing values (denoted as JMI$^{aux}$) (36).

All imputation methods can be directly applied to individuals and only require access to estimated population characteristics (i.e. mean and covariance estimates of the predictors) to account for missing predictor values. For both imputation methods the required population characteristics are easily stored and accessible in 'live' clinical practice within any accompanying CDSS. The outcome is excluded from the imputation procedure as this information is not available when imputing the missing predictor values, and is the target of the prediction model. The corresponding source code is available from the supplementary information (Appendix E).

**2.3  Study population**

The UCC-CVRM sample consisted of 3.880 patients with 23 variables and the UCC-SMART study consisted of 12.616 patients with 155 variables. Some patient values were missing in UCC-CVRM (for 1057/3880 patients) and in UCC-SMART (for 2028/12616 patients). For the purpose of our methodological study we had to have complete control over the patterns of missing predictor data and the true underlying predictor values, and needed to start with a fully observed data set that could be considered as the reference situation. To that end, for each dataset separately, we imputed all missing data once using Multiple Imputation by Chained Equations (for UCC-SMART) and nearest neighbor imputation (for UCC-CVRM) (34). These then completed data sets formed the reference situation after which missing predictor values were generated according to various patterns (see below). Table 1 provides an overview of the completed variables in both cohorts, and how they were subsequently used in our simulation study. To assess the relatedness between UCC-CVRM and UCC-SMART, we calculated the membership c-statistic (37), which ranges between 0.5 (both samples have a similar case-mix) and 1 (the case-mix between both samples does not have any overlap). We found a membership c of 0.86, which indicates that the population characteristics of UCC-CVRM and UCC-SMART differ greatly.

**2.4  Simulation study**

We performed 4 simulation studies to investigate the impact of real-time predictor imputation on absolute risk predictions (Figure 1). In the first simulation, we considered the ideal situation where a (new) patient stems from the same population (i.e. UCC-SMART) as the one that is used to develop the prediction model, to derive the population characteristics, and to test the accuracy of individual risk predictions after the real-time imputations. In the second simulation, we considered a less ideal situation where imputations are based on the characteristics from a different, but related, population (i.e. UCC-CVRM). This simulation mimics the situation where development data are unavailable (or otherwise insufficient) to inform the imputation procedure, and thus assesses the transportability of the imputation model. In the third simulation, we investigated the situation where the estimated population characteristics underlying the imputations are derived from an external cohort (UCC-CVRM) and subsequently updated using local data (from UCC-SMART). This resembles a situation in which a small amount of local data is available, though insufficient to entirely inform the real-time imputation procedure. In the final simulation, we considered the most extreme scenario where 3 different populations are used to derive a prediction model (Framingham Risk Score (9)), the imputation model (UCC-CVRM), and to test the accuracy of the real-time imputations on the individuals' absolute risk predictions (UCC-SMART). This simulation mimics a more common predicament in which local data is insufficient to inform the imputation procedure and there is no access to the data from which the prediction model had been developed.

In all simulation studies, we considered UCC-SMART as the target population. For simulations 1-3, we adopted a leave-one-out-cross-validation (LOOCV) approach to develop the prediction model, to derive the population characteristics, and to evaluate the accuracy of risk predictions. This procedure ensures that independent data are used for the evaluation of risk predictions. In the LOOCV approach both the prediction model imputation model were derived from all but one patient (leave-one-out) of UCC-SMART. In the remaining hold-out patient, one or more predictor variables were then set to missing (see Figure 2 for an overview of which sets of predictor values were set to missing). The leave-one-out procedure was repeated until all patients had been removed from UCC-SMART exactly once (Figure 3).

LOOCV was not needed for the 4th simulation as each task (prediction model development, derivation of population characteristics, and evaluation of risk predictions) involved a different dataset (4).

### 2.4.1 Step 1. Estimation of the prediction model

For all simulation studies, the prediction model of interest was a Cox proportional hazards model predicting the onset of cardiovascular disease or coronary death. This model was derived in the LOO (leave-one-out) subset of UCC-SMART using predictors from the original FRS (simulation 1-3), or retrieved from the literature (simulation 4). A detailed description of how the prediction models were fit and the R code is listed in Appendix E. As a sensitivity analysis, we fitted a Cox regression model with only age and gender as predictors and included a scenario where, though unrealistic, age and gender were missing.

### 2.4.2 Step 2: Estimation of the population characteristics

We estimated the population characteristics necessary for the real-time missing data methods (i.e. the imputation models) in the following data (Figure 1):

- in the entire LOO subset of UCC-SMART (simulation 1),
- in the entire dataset of UCC-CVRM (simulation 2 and 4)
- in the entire dataset of UCC-CVRM, plus a random sample of the LOO subset of UCC-SMART, which were simply stacked. (simulation 3)

*2.4.3    Step 3: Introduction and imputation of missing values*

For simulation 1-3, we set one or more predictor variables to missing in each hold-out patient of UCC-SMART (scenarios illustrated in Figure 2). To match the introduction of missing values with real life occurrences of missingness, we included scenarios based on observed patterns of missingness in UCC-CVRM. For simulation 4, missing values were generated for the entire UCC-SMART dataset, rather than for individual patients. We subsequently impute the missing values once using the following strategies:

1. Mean imputation. Any missing predictor value was imputed with their respective mean as estimated in step 2.

2. JMI with observed predictors only. Each missing predictor value is replaced by its expected value conditional on the individual's observed *predictors*. The expected value is derived using the estimated population means and covariances from step 2.

3. JMI with observed predictors and auxiliary variables. Each missing predictor value is replaced by its expected values conditional on *all* the observed patient data. Note that this includes additional patient data that are not included as predictors in the prediction model (Table 1).

*2.4.4    Step 4 – Risk prediction and validation of model performance*

The imputed missing predictor values were then used together with the observed predictor values to calculate the linear predictor $\eta_i$ (where $\eta_i = \beta_1 x_{i1} + \beta_2 x_{i2} + ...$) and the 10-year predicted absolute risk. The predictions from all UCC-SMART patients were then used to assess the following performance measures: 1) Mean Squared Error (MSE) of the prediction model's linear predictor, (2) concordance (C-)statistic, 3) calibration-in-the-large, 4) the calibration slope and 5) the decision curve (4,16,38,39).

1) *The MSE of the linear predictor of the prediction model* can be described as the average squared difference between the linear predictor after imputation and the true, original linear predictor (i.e. before introducing missing values) (40). The linear predictor can be described as the weighted sum of the predictors of a given patient, where the weights consist of the model coefficients (38). Lower values for the MSE are preferred.

2) *The C-statistic* can be described as the ability of the model to discern those who have experienced an event and those who haven't (7,40,41). It is represented by the probability of correctly discerning who, between two random subjects, has the higher predicted probability of survival. The C-statistic is ideally close to 1.

3) *Calibration-in-the-large (CITL)* can be described as the overall calibration of the model (i.e. agreement between average predicted risk and average observed risk) (4,7,41,42). It is interpreted as an indication of the extent to which the predictions systematically over- or underestimate the risk; the ideal value is 0.

4) *The calibration slope* can be described as a quantification of the extent that predicted risks vary too much (slope <1) or too little (slope > 1), and is often used as an indication of overfitting or lack of transportability (7,16,40–42). The ideal value is 1.

5) *The decision curve* can be described as a way of identifying the potential impact of leveraging individual risk predictions for decision making (4,39,43). It considers a range of thresholds (e.g. 10%) to classify patients into high risk (indication of treatment) or low risk (no treatment required) and calculates the net benefit (NB) for each cut-off value. A decision curve is then constructed for 3 different treatment strategies: treat all, treat none, or treat according to risk predictions. Ideally, the decision curve of the latter strategy depicts consistently better NB over the complete range of thresholds.

# 3 Results

## 3.1 Prediction model performance in the absence of missing values

Based on internal validation by means of LOOCV, the optimism corrected c-statistic for our newly derived prediction model in UCC-SMART was 0.705. As expected, the CITL and calibration slope were near 0 (-0.0005) and 1 (0.9999) respectively. Therefore, there were no signs of miscalibrations and/or over/underfitting of the developed CVD risk prediction model. The prediction model that was based on age and gender yielded an optimism corrected c-statistic of 0.679, with a slope of 0.9999 and an intercept of -0.00005. Finally, the refitted FRS model (as derived from the literature) yielded a c-statistic of 0.6280 and a slope of 0.8205 in UCC-SMART.

## *3.2* **Prediction model performance in presence of missing data**

### *3.2.1 Mean squared error*

The MSE of the linear predictor was consistently lower when adopting JMI, as compared to M-Imp. The implementation of JMI was particularly advantageous when adjusting for auxiliary variables that were not part of the prediction model (see table 2 for the results of scenario 1 and 5). For instance, when total cholesterol (TC), HDL-cholesterol (HDL-c), use of Antihypertensive Drugs (AD), smoking and Diabetes Mellitus (DM) were missing (i.e. scenario 5), M-Imp yielded an MSE of 0.130, whereas the MSE for JMI was 0.126 or even 0.101 when utilizing auxiliary variables. As expected, differences in MSE were lower, when imputing other predictors that did not have a strong contribution in the prediction model, or much more pronounced when imputing important predictors (see table 3 for the results of the sensitivity analysis with age and gender missing). This expected discrepancy results from the fact that the linear predictor is a weighted average of the predictors and the important variables simply have larger weights. When imputation was based on the characteristics of a different, but related, cohort to UCC-SMART, all imputation strategies yielded a substantially larger MSE. For instance, when TC, HDL-c, AD, smoking and DM were missing (i.e. scenario 5), the MSE increased from 0.130 to 0.193 for M-Imp, and from 0.1014 to 0.159 for JMI$^{aux}$. Again, JMI$^{aux}$ was superior to M-Imp and JMI based on predictor variables only. As expected, the MSE for all imputation methods improved when the imputation model was based on a mixture of patients from both the UCC-CVRM (different but related) and the UCC-SMART (the target cohort for predictions). However, the lowest MSE's were obtained when imputations were based on UCC-SMART data only.

*3.2.2    C-statistic*

The c-statistic was higher for both implementations of JMI, when compared to M-Imp (Table 2). Using JMI$^{aux}$ further increased the c-statistic substantially, especially when important predictors (i.e. age and gender) were missing (Table 3). In this scenario, M-Imp yielded a c-statistic of 0.61, whereas JMI yielded a c-statistic of 0.62 or even 0.67 if auxiliary variables were used. Discrimination performance did not much deteriorate when imputation was based on the characteristics from a different but related population. Again, JMI$^{aux}$ was superior to M-Imp and JMI based on predictor variables only. The c-statistic, for all imputation methods, improved when the population characteristics from UCC-CVRM were augmented with data from UCC-SMART. However, when an external prediction model was used in combination with external population characteristics (simulation 4), the utilization of auxiliary variables did not seem to improve on the discriminatory ability of risk predictions (Table 4). The highest c-statistics were obtained when imputations were based on UCC-SMART data only and a locally derived prediction model was used.

*3.2.3    Calibration-in-the-large*

The CITL was consistently closer to the ideal value (i.e. 0) for all scenarios when using both implementations of JMI, when compared to M-Imp. Using JMI$^{aux}$ improved the CITLs further towards their ideal value (Table 2). When imputation used estimated population characteristics from UCC-CVRM, all imputation strategies had a substantially worse CITL. The performance drop was most notable as more predictors in the model were missing. Again, JMI$^{aux}$ was superior to M-Imp and JMI based on predictor variables only. The CITL, for all imputation methods, improved when the population characteristics from UCC-CVRM were augmented with data from UCC-SMART. When an external prediction model was used, M-Imp yielded the "best" CITL (-0.167 as opposed to -0.2030 for JMI and -0.2256 for JMI$^{aux}$; Table 4). The CITLs were closest to 0 when imputations were based on UCC-SMART data only.

### 3.2.4 Calibration slope

The use of JMI$^{aux}$ improved the calibration slope as compared to M-Imp or JMI using predictor variables only (Table 2). When imputation used population characteristics from UCC-CVRM, the variability of predicted risks generally became too large (slope < 1 for all imputation methods). The performance drop was most notable as more predictors were missing. When an external prediction model was used, both JMI and JMI$^{aux}$ yielded better calibration as compared to M-Imp (Table 4), although JMI$^{aux}$ performed worse than JMI. The best calibration slopes were found for imputations based on UCC-SMART data only.

Figure 5 visualizes calibration plots for scenarios 1, 5 and 8. It shows that when important predictors (i.e. age and gender in scenario 8) are missing there is a notable impact on the calibration of 10-year risk predictions, especially when using external data for generating imputations. When less important predictors are missing (scenario 1 and 5) the differences between the imputation methods are much less pronounced in the calibration plots.

### 3.2.5 Decision curve

When important variables were missing, imputation through JMI with auxiliary variables yielded an improved net benefit over the whole range of thresholds when compared to M-Imp and JMI (Figure 6), and was substantially better than treat-all or treat-none strategies. The observed net benefit did not much deteriorate when imputation was based on a different, but related, dataset.

A complete detailed overview of all results (e.g. all scenarios) can be found in the supplementary material.

## 4   Discussion

Our aim was to evaluate the impact of using real-time imputation of missing predictor values on the performance of cardiovascular risk prediction models in individual patients. We considered mean imputation and joint modeling imputation to provide automated real-time imputations. Our results demonstrate that in all scenarios and for all parameters studied (c-index, calibration and decision curve analysis) JMI leads to more accurate risk predictions than M-Imp, especially when used to impute a higher number of missing predictors (e.g. scenario 5 for prediction of cardiovascular events). The performance of JMI greatly improved when imputations were based on all observed patient data, and not restricted to only the predictors that were in the prediction model. Finally, we found that real-time missing predictor imputations were most accurate when the imputation method relied on characteristics that were directly estimated a sample from the target population (i.e. the population for which predictions are required), rather than from an external though related dataset. In the latter case, while discriminative performance was stable, calibration clearly deteriorated (in terms of both CITL and calibration slope). This implies that the need for local updating, as is well known in clinical prediction modeling, may extend to imputation models. In practice, a prediction model is ideally developed together with an appropriate missing data method for real-time imputation. When high quality local data are available, performance gains can be expected for that setting by local updating of both the prediction model and the imputation model.

Our findings suggest that JMI should be preferred over M-Imp for real-time imputation of missing predictor values in routine care, ideally making use of additional patient data (variables) that are not part of the prediction model. The underlying rationale, is that some variables that are highly correlated are unlikely to both end up in a prediction model (due to little added value), but are quite valuable for imputation purposes when one or the other is missing. The implementation of JMI is very straightforward, and only requires estimating the mean and covariance of all relevant patient variables in a representative sample. Imputations are then generated using a set of mathematical equations that are well established in the statistical literature (23). As JMI does not rely on disease-specific patient characteristic and lends itself excellently for local tailoring (44), it is considered highly scalable to a multitude of clinical settings and populations. Routine reporting of population characteristics (i.e. means and covariance) would greatly facilitate the implementation of risk prediction models in the presence of missing predictor data in daily practice, and has previously been recommended to improve the interpretation of validation study results (37).

A limitation we observed in the data was that most of the explained variability in risk of cardiovascular disease, as defined in our study, could be inferred based on age and gender. Although additional predictors (e.g. blood pressure, cholesterol levels) somewhat improved the model's discrimination and calibration performance, their individual added value appears small. A further limitation of the data was the lack of strong correlations between predictors other than age and gender (appendix D). Consequently, the information available for JMI to leverage observed patient characteristics was limited. These findings are in line with earlier research, suggesting that M-Imp performs similarly to more advanced imputation methods when considering commonly encountered missing data patterns in cardiovascular routine care (45). However, our study reveals that JMI had the advantage even under these typical but difficult settings. Gains are expected to be larger when the interrelation of predictors is stronger and especially when key auxiliary variables can be identified. Moreover, for many disease areas, risk prediction relies more strongly on a multitude patient characteristic that are more likely to be missing (e.g. certain imaging characteristics, biomarkers or genetic profiles), and JMI offers a larger advantage.

Various other aspects need to be addressed to fully appreciate these results. First, we restricted our comparison to M-Imp and JMI. Considering M-Imp was picked as a comparator, we choose JMI as it was well established in the statistical literature and permitted relatively straightforward adjustments to be applied in clinical practice via the EHR (29,34). Other, more flexible, imputation strategies exist, and have been discussed at length (23). These strategies generally require more complex descriptions of the population characteristics and adopt more advanced procedures to generate imputations. For this reason, their implementation appears less straightforward in routine care. A more detailed overview of the impact of using other strategies for handling real-time missing predictor value imputation is warranted. Also, the use of multiple imputation may be preferable with respect to prediction accuracy in case of models with a non-linear link function such as the Cox or logistic model, the reason is multiple imputation can correctly convey the influence of imputation uncertainty on the expected prediction. The available R code already provides in this, though in this study we explicitly choose to use single imputation. We choose single imputation due to its convenience in real-time clinical practice. The imputation process is quick, in contrast to the usually computationally expensive multiple imputation, and it presents an individual's imputed predictor value which may be informative to the clinician. Additionally, rather than imputing a random draw, we impute the most likely value in order to be able to easily reproduce model predictions from the imputed data. Ideally, the predictions would be based on multiple imputation from the conditional distribution of the missing predictors rather than representing their conditional means. Further extensions, for example multilevel multiple imputation, may also be recommended in specific situations where the prediction model and accompanying imputation models are derived from large datasets with clustering (46). Lastly, whilst there are many clinical settings and populations the study only considered cardiovascular risk prediction. The performance of JMI, when compared to M-Imp, might have been further emphasized had other clinical settings been considered.

In summary, this study evaluates the use of two imputation methods for handling missing predictor values when applying risk prediction models in daily practice. We recommend JMI over mean imputation, preferably based on estimated from local data and with the use of available auxiliary variables. The added value of JMI is most evident when missing predictors are associated with either observed predictor values or auxiliary variables.

# 5 Disclosures

## 5.1 Funding sources

This work was supported by the Netherlands Heart foundation (public-private study grant, grant number: #2018B006); and the Top Sector Life Sciences & health (PPP allowance made available to Netherlands Heart Foundation to stimulate public-private partnerships).

## 5.2 Conflicts of interest

None declared.

## 5.3 Author contributions

TD, KM, KG, and SN conceived of the presented idea, in correspondence with earlier work by FA, FV, MB and KG. SN, TD and JH derived the models and analyzed the data. TD and JH verified the analytical methods. TD supervised the findings of this work. SN, KG, TD and JH contributed to the interpretation of the results. SN wrote the manuscript. All authors provided critical feedback and helped shape the research, analysis, and manuscript.

## 5.4 Data availability statement

The data that support the findings of this study are available from the UCC upon reasonable request (https://www.umcutrecht.nl/en/Research/Strategic-themes/Circulatory-Health/Facilities/UCC).

## 5.5 Acknowledgements

We gratefully acknowledge the contribution of the research nurses; R. van Petersen (data-manager); B. van Dinther (study manager) and the members of the Utrecht Cardiovascular Cohort-Second Manifestations of ARTerial disease-Studygroup (UCC-SMART-Studygroup): F.W.

Asselbergs and H.M. Nathoe, Department of Cardiology; G.J. de Borst, Department of Vascular Surgery; M.L. Bots and M.I. Geerlings, Julius Center for health Sciences and Primary Care; M.H. Emmelot, Department of Geriatrics; P.A. de Jong and T. Leiner, Department of Radiology; A.T. Lely, Department of Obstetrics & Gynecology; N.P. van der Kaaij, Department of Cardiothoracic Surgery; L.J. Kappelle and Y.M. Ruigrok, Department of Neurology; M.C. Verhaar, Department of Nephrology, F.L.J. Visseren (chair) and J. Westerink, Department of Vascular Medicine, University Medical Center Utrecht and Utrecht University.

**Box 1: An example of real-time imputation in an individual patient**

> *Example. A patient visits their physician for a regular check-up. The patient and physician have access to a clinical decision support system that provides information on previously ordered test results (automatically retrieved from a registry). The physician would like to know the 10-year risk for the patient to suffer from a cardiovascular event, in order to determine whether any lifestyle changes or preventative therapies are needed. A calculator to determine this risk (e.g. the pooled cohort equations) is incorporated in the clinical decision support system, but requires complete information on several patient characteristics, including their BMI, cholesterol levels, and blood pressure. Many of these predictors are directly available (e.g. age, gender) at the visit. However, for some patients, important lab results (e.g. LDL cholesterol) are yet unknown or outdated (e.g. when retrieved from the registry). It is then not possible to determine the absolute risk of CVD for these patients. Our algorithm provides a substitute value for the missing LDL-cholesterol in real-time, enabling the calculation of a risk estimate 'on the spot'.*

**Table 1: general characteristics of the study populations**

|  | **UCC-SMART** Mean (sd) or n/total (%)* | Role | **UCC-CVRM** Mean (sd) or n/total (%)** | Role |
|---|---|---|---|---|
| Age (years) | 56.28 (12.45) | Predictor | 61.7 (18.18) | Predictor |
| Gender (1=male) | 8258 (65.50) | Predictor | 1987 (51.21) | Predictor |
| Smoking (1=yes) | 3560 (28.24) | Predictor | 363 (9.36) | Predictor |
| SBP (mmHg) | 144.67 (21.58) | Predictor | 142.75 (24.24) | Predictor |
| TC (mmol/l) | 5.11 (1.37) | Predictor | 5.07 (1.24) | Predictor |
| HDL-c (mmol/l) | 1.27 (0.38) | Predictor | 1.36 (0.36) | Predictor |
| DM (1=yes) | 2299 (18.23) | Predictor | 755 (19.46) | Predictor |
| AD (1=yes) | 8332 (66.09) | Predictor | 705 (18.17) | Predictor |
| LDL-c (mmol/l) | 3.15 (1.22) | auxiliary | 3.08 (1.27) | auxiliary |
| HbA1c (mmol/mol) | 3.69 (0.20) | auxiliary | 3.66 (0.22) | auxiliary |
| MDRD (ml/min/1.73m2) | 79.90 (19.54) | auxiliary | 81.79 (24.56) | auxiliary |
| History of CVD (1=yes) | 8134 (64.51) | auxiliary | 1971 (50.80) | auxiliary |
| Time since 1st CVD event (years) | 2.37 (5.93) | auxiliary | 4.642 (8.06) | auxiliary |
| MPKR (mg/mmol) | 4.10 (13.71) | auxiliary | NA | None |
| CRP (mg/L) | 0.71 (1.13) | auxiliary | NA | None |
| AF (1=yes) | 164 (1.30) | auxiliary | NA | None |
| LLD (1=yes) | 6836 (54.22%) | auxiliary | NA | None |
| PAI (1=yes) | 6805 (53.97%) | auxiliary | NA | None |

Legend – SBP: systolic blood pressure, TC: total cholesterol, HDL-c: high-density lipoprotein cholesterol, , DM: diabetes mellitus, AD: antihypertensive drugs, LDL-c: low-density lipoprotein cholesterol, HbA1c: glycated hemoglobin, MDRD: modification of diet in renal diseases, MPKR: micro-protein/creatinine ratio, AF: atrial fibrillation, lipid-lowering drugs, PAI: platelet aggregation inhibitors. * after multiple imputation by chained equations.

**Table 2. Results of simulating scenarios with small and large amounts of missing data**

| Scenario 1 (small amount of missing data): (1) SBP, (2) smoking | Imputation methods | MSE of the LP (% difference to M-Imp) | C-index | CITL | Calibration slope |
|---|---|---|---|---|---|
| *Apparent performance (reference)* | | | 0.7051 | -0.0001 | 0.9999 |
| **Simulation 1** Local data (for informing imputation) | M-Imp | 0.0702 | 0.6908 | 0.0228 | 0.9415 |
| | JMI | 0.0685 (-2.35%) | 0.6913 | 0.0242 | 0.9552 |
| | JMI$^{aux}$ | 0.0649 (-7.50%) | 0.6975 | 0.0221 | 0.9928 |
| **Simulation 2** External data (for informing imputation) | M-Imp | 0.0802 | 0.6908 | 0.1227 | 0.9415 |
| | JMI | 0.0782 (-2.56%) | 0.6911 | 0.1018 | 0.9269 |
| | JMI$^{aux}$ | 0.0801 (0.001%) | 0.6902 | 0.1123 | 0.9251 |
| **Simulation 3** External data with 1.500 local patients | M-Imp | 0.0746 | 0.6909 | 0.0845 | 0.9393 |
| | JMI | 0.0718 (-3.90%) | 0.6913 | 0.0510 | 0.9278 |
| | JMI$^{aux}$ | 0.0708 (-5.37%) | 0.6911 | 0.0485 | 0.9315 |

| Scenario 5 (large amount of missing data): (1) TC, (2) HDL-c, (3) AD (4) smoking, (5) DM missing | Imputation methods | MSE of the LP (% difference to M-Imp) | C-index | CITL | Calibration slope |
|---|---|---|---|---|---|
| *Apparent performance (reference)* | | | 0.7051 | -0.0001 | 0.9999 |
| **Simulation 1** Local data (for informing imputation) | M-Imp | 0.1300 | 0.6797 | 0.0581 | 0.9199 |
| | JMI | 0.1262 (-2.98%) | 0.6803 | 0.0549 | 0.9211 |
| | JMI$^{aux}$ | 0.1014 (-21.98%) | 0.6960 | 0.0369 | 1.0052 |
| **Simulation 2** External data (for informing imputation) | M-Imp | 0.1930 | 0.6797 | 0.3090 | 0.9199 |
| | JMI | 0.1806 (-6.42%) | 0.6803 | 0.2797 | 0.9067 |
| | JMI$^{aux}$ | 0.1591 (-17.57%) | 0.6844 | 0.2595 | 0.9475 |
| **Simulation 3** External data with 1.500 local patients | M-Imp | 0.1683 | 0.6790 | 0.2418 | 0.9387 |
| | JMI | 0.1603 (-4.78%) | 0.6792 | 0.2078 | 0.9095 |
| | JMI$^{aux}$ | 0.1334 (-20.72%) | 0.6851 | 0.1677 | 0.9573 |

Legend – SBP: systolic blood pressure, TC: total cholesterol, HDL-c: HDL-cholesterol, AD: antihypertensive drug, SBP: systolic blood pressure, DM: diabetes mellitus, MSE: mean squared error, LP: linear predictor, CITL: calibration in the large, M-Imp: mean imputation, JMI: joint modelling imputation, JMI$^{aux}$: joint modelling imputation with auxiliary variables.

**Table 3. Results sensitivity analysis**

| Scenario 8:<br>(1) Age, (2) gender missing | Imputation methods | MSE of the LP<br>(% difference to M-Imp) | C-index | CITL | Calibration slope |
|---|---|---|---|---|---|
| *Apparent performance (reference)* | | | 0.7051 | -0.0001 | 0.9999 |
| **Simulation 1**<br>Local data (for informing imputation) | M-Imp | 0.7438 | 0.6063 | 0.1958 | 0.8225 |
| | JMI | 0.6373 (-14.32%) | 0.6223 | 0.1616 | 0.8052 |
| | JMI$^{aux}$ | 0.4517 (-39.26%) | 0.6931 | 0.0794 | 1.0828 |
| **Simulation 2**<br>External data (for informing imputation) | M-Imp | 0.8334 | 0.6064 | -0.1037 | 0.8230 |
| | JMI | 0.7963 (-4.45%) | 0.6116 | -0.2221 | 0.5769 |
| | JMI$^{aux}$ | 0.7018 (-15.79%) | 0.6721 | -0.3649 | 0.8453 |
| **Simulation 3**<br>External data with 1.500 local patients | M-Imp | 0.792383 | 0.6107 | -0.0205 | 0.8429 |
| | JMI | 0.7252996 (-9.25%) | 0.6131 | -0.0659 | 0.6480 |
| | JMI$^{aux}$ | 0.5739753 (-38.05% | 0.6856 | -0.1451 | 0.9654 |

Legend – MSE: mean squared error, LP: linear predictor, CITL: calibration in the large, M-Imp: mean imputation, JMI: joint modelling imputation, JMI+: joint modelling imputation with auxiliary variables.

**Table 4. Multivariable missing data imputation (simulation 4): the use of an external prediction and imputation model**

| Combination of all missing data scenarios | Imputation methods | MSE of the LP<br>(% difference to M-Imp) | C-index | CITL | Calibration slope |
|---|---|---|---|---|---|
| *Reference when no variables are missing* | | | 0.6280 | -0.0888 | 0.8205 |
| **Simulation 4**<br>External prediction and imputation model | M-Imp | 0.1689 | 0.6095 | -0.1674 | 0.7424 |
| | JMI | 0.1585 (-6.56%) | 0.6145 | -0.2030 | 0.7549 |
| | JMI$^{aux}$ | 0.1334 (-26.61%) | 0.6135 | -0.2257 | 0.7495 |

Legend – CITL: calibration in the large, M-Imp: mean imputation, JMI: joint modelling imputation, JMI$^{aux}$: joint modelling imputation with auxiliary variables.

**Figure 1: the simulation studies illustrated**

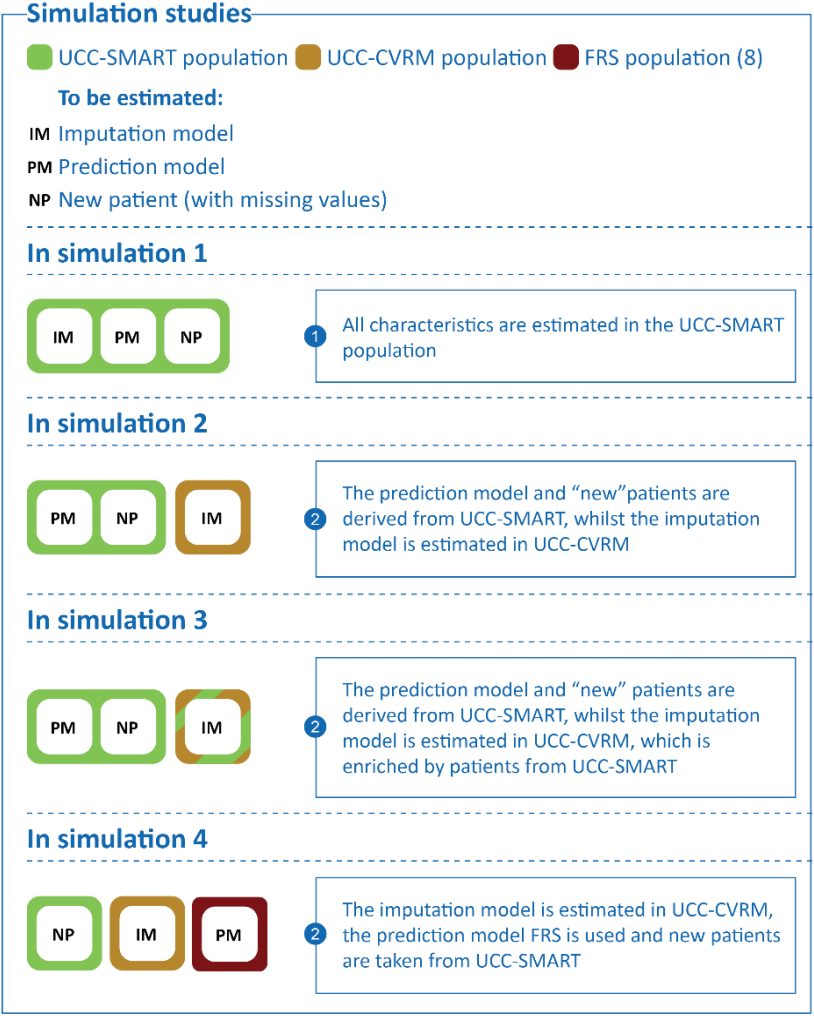

**Figure 2: Multivariate scenarios of missing predictor values observed in UCC-CVRM**

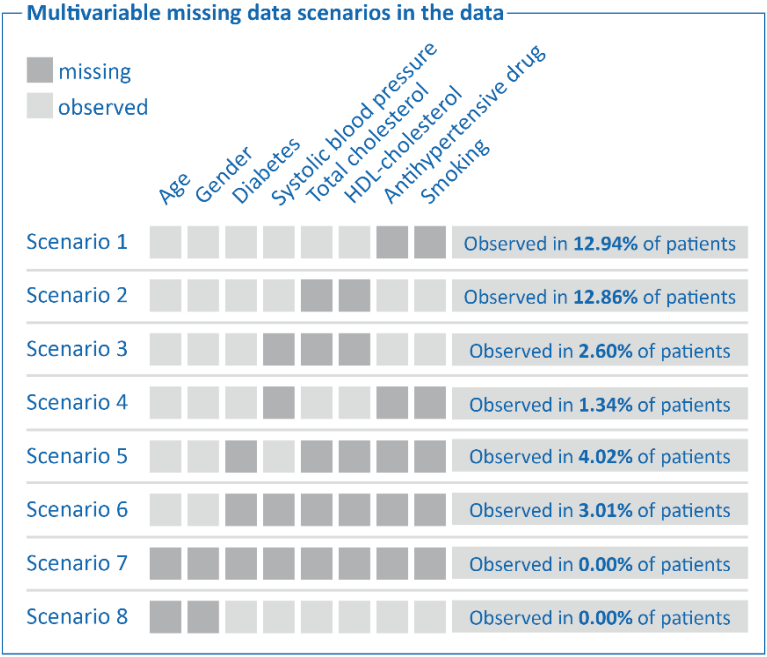

## Figure 3: Simulation study 1-3 in detail

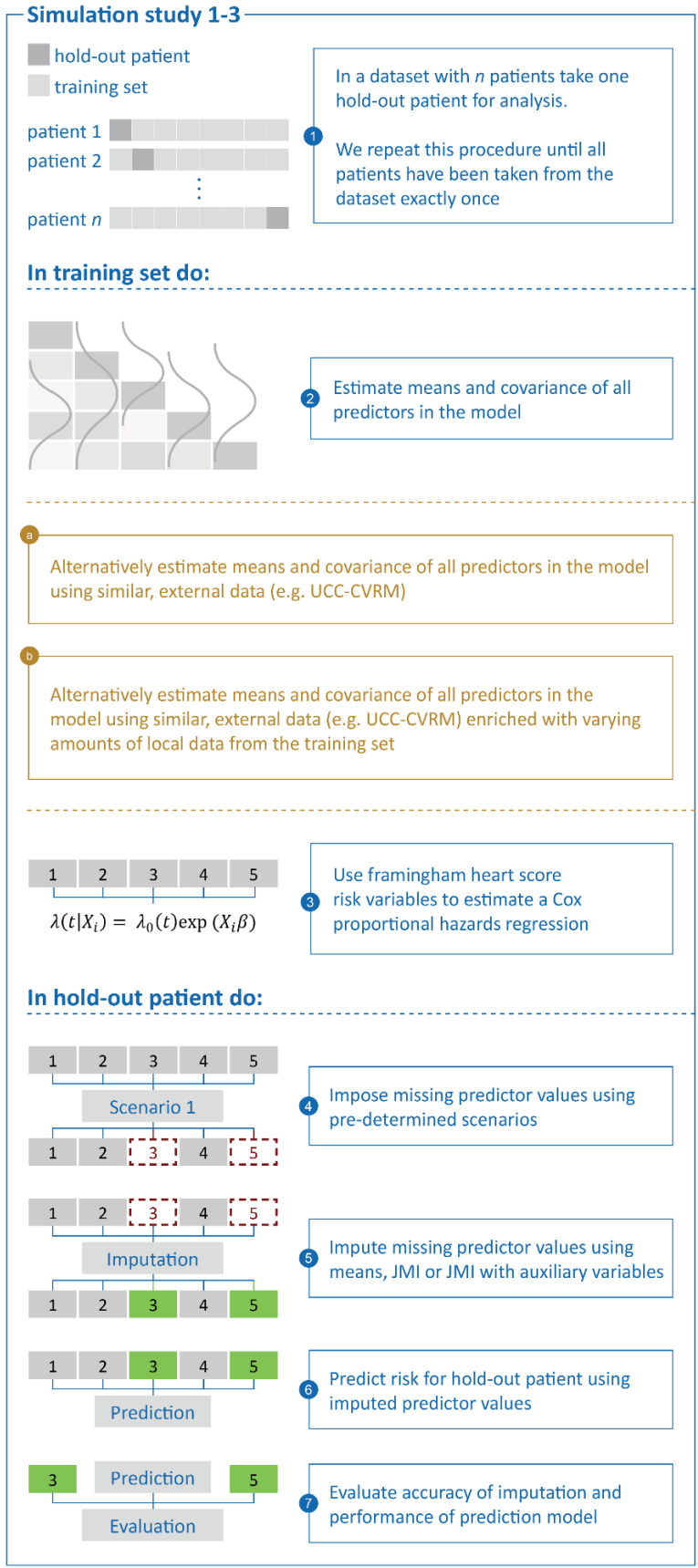

**Figure 4: Simulation study 4 in detail**

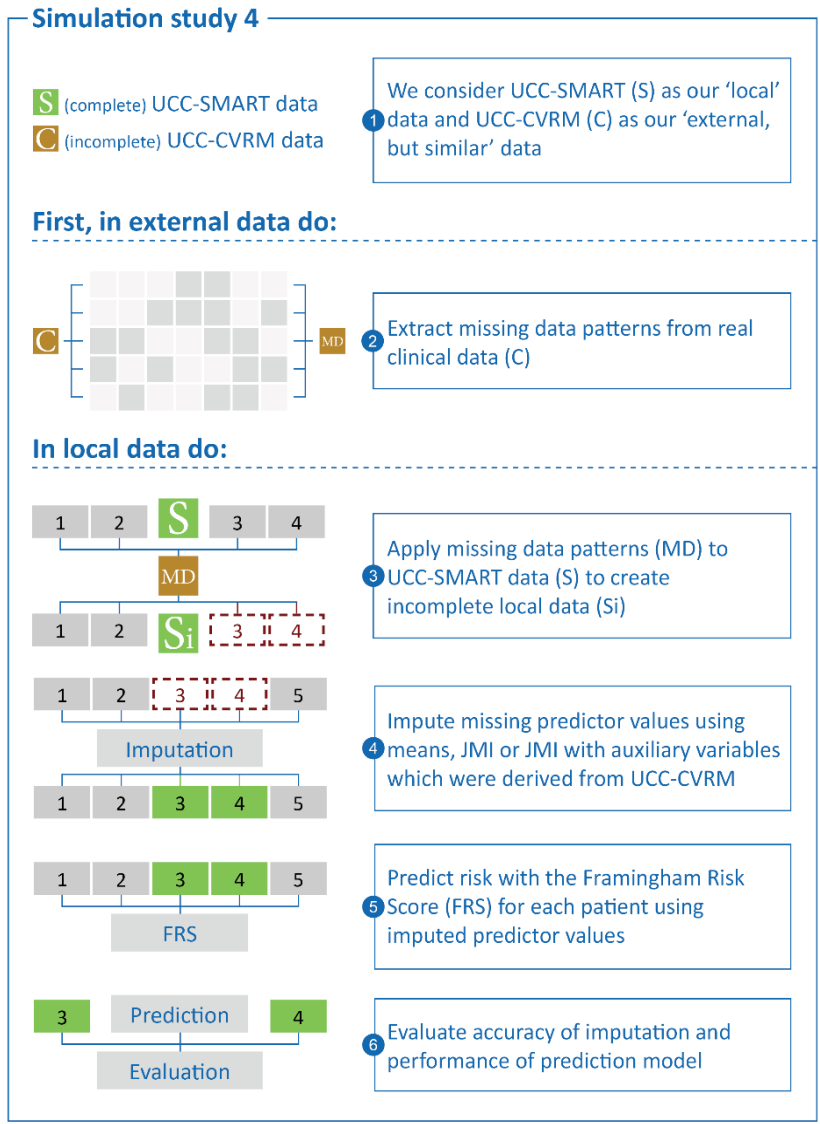

**Figure 5: Calibration plots for scenario 1, 5 and 8**

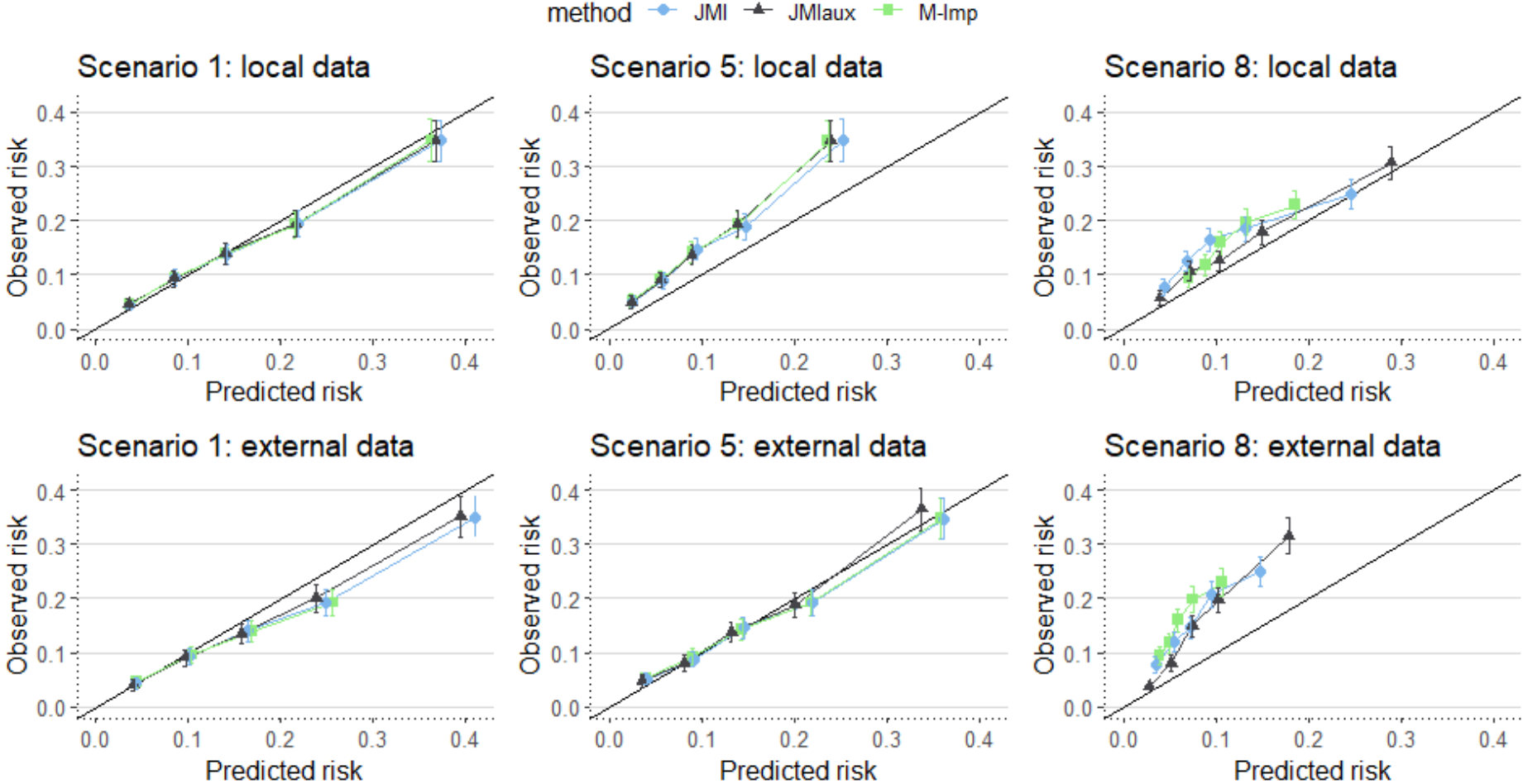

# Figure 6: Decision curve analysis simulation 1

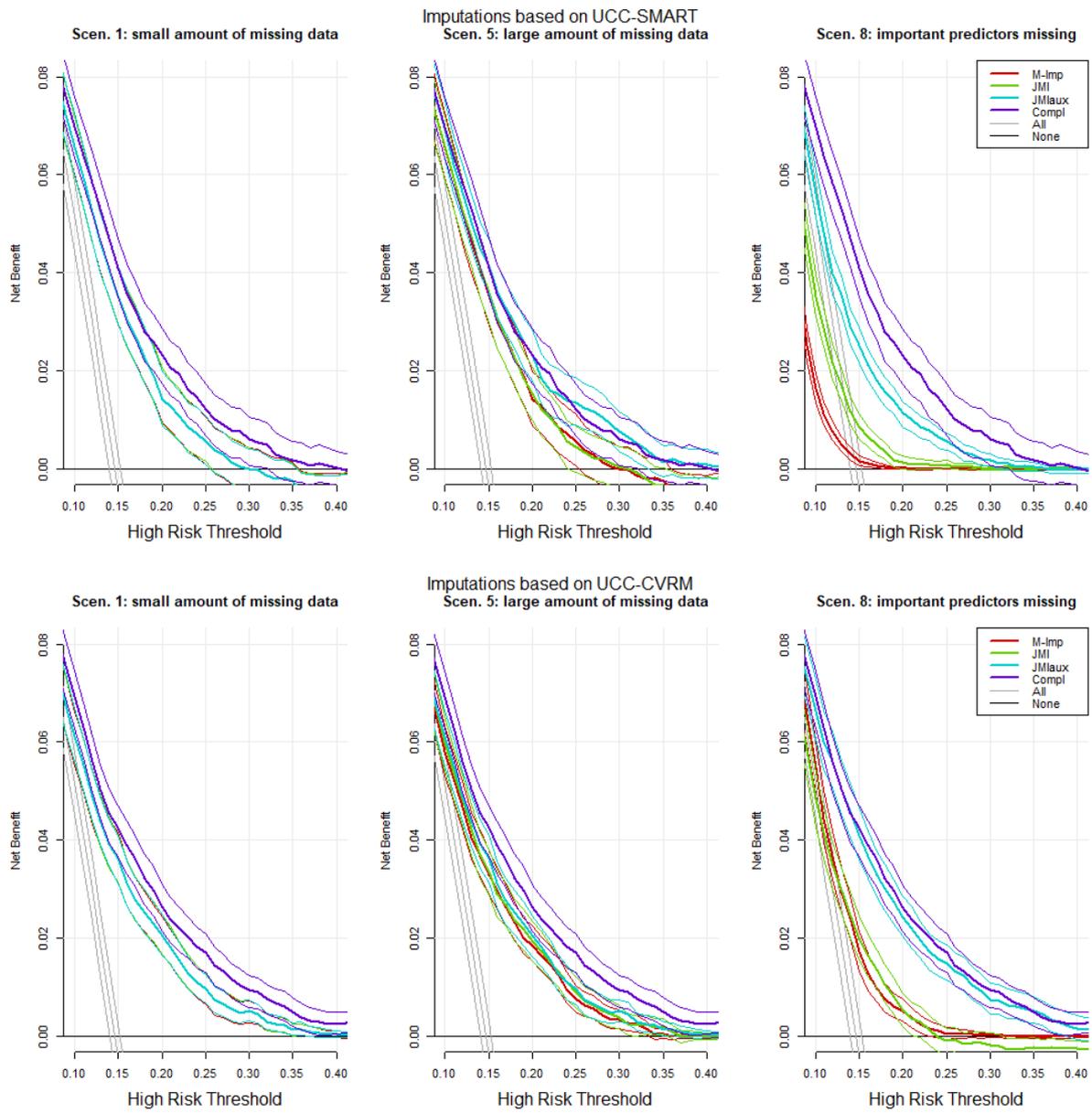

**Appendix A simulation 1: imputing multiple missing predictor values scenarios using local data**

| # | Variables missing | Methods | MSE of the LP (% difference to M-Imp) | C-index | CITL | Calibration slope |
|---|---|---|---|---|---|---|
| | *Apparent performance (reference)* | | | 0.7051 | -0.0001 | 0.9999 |
| 1 | (1) SBP, (2) smoking | M-Imp | 0.0702 | 0.6908 | 0.0228 | 0.9415 |
| | | JMI | 0.0685 (-2.35%) | 0.6913 | 0.0242 | 0.9552 |
| | | JMI$^{aux}$ | 0.0649 (-7.50%) | 0.6975 | 0.0221 | 0.9928 |
| 2 | (1) TC, (2) HDL-c | M-Imp | 0.0265 | 0.7005 | 0.0230 | 0.9766 |
| | | JMI | 0.0241 (-8.97%) | 0.7013 | 0.0192 | 0.9751 |
| | | JMI$^{aux}$ | 0.0220 (-16.84%) | 0.7046 | 0.0153 | 0.9901 |
| 3 | (1) TC, (2) HDL-c, (3) SBP | M-Imp | 0.0333 | 0.6994 | 0.0406 | 1.0003 |
| | | JMI | 0.0314 (-5.74%) | 0.7000 | 0.0268 | 0.9787 |
| | | JMI$^{aux}$ | 0.0276 (-17.26%) | 0.7040 | 0.0212 | 0.9935 |
| 4 | (1) TC, (2) HDL-c, (3) SBP, (4) AD | M-Imp | 0.0459 | 0.6981 | 0.0546 | 1.0217 |
| | | JMI | 0.0425 (-7.41%) | 0.6983 | 0.0255 | 0.9702 |
| | | JMI$^{aux}$ | 0.0394 (-14.13%) | 0.7044 | 0.0231 | 1.0010 |
| 5 | (1) TC, (2) HDL-c, (3) AD (4) smoking, (5) DM | M-Imp | 0.1300 | 0.6797 | 0.0581 | 0.9199 |
| | | JMI | 0.1262 (-2.98%) | 0.6803 | 0.0549 | 0.9211 |
| | | JMI$^{aux}$ | 0.1014 (-21.98%) | 0.6960 | 0.0369 | 1.0052 |
| 6 | (1) TC, (2) HDL-c, (3) AD, (4) smoking, (5) DM, (6) SBP | M-Imp | 0.1383 | 0.6785 | 0.0758 | 0.9430 |
| | | JMI | 0.1351 (-2.31%) | 0.6788 | 0.0637 | 0.9249 |
| | | JMI$^{aux}$ | 0.1087 (-21.45%) | 0.6955 | 0.0441 | 1.0128 |
| 7 | (1) Age, (2) gender, (3) TC, (4) HDL-c, (5) AD, (6) smoking, (7) DM, (8) SBP | M-Imp | 0.9137 | 0.5112 | 0.2897 | 77.819 |
| | | JMI | 0.9137 (0.00%) | 0.5112 | 0.2897 | 77.819 |
| | | JMI$^{aux}$ | 0.5990 (-34.44%) | 0.6892 | 0.1544 | 1.3754 |
| 8 | (1) Age, (2) gender | M-Imp | 0.7438 | 0.6063 | 0.1958 | 0.8225 |
| | | JMI | 0.6373 (-14.32%) | 0.6223 | 0.1616 | 0.8052 |
| | | JMI$^{aux}$ | 0.4517 (-39.26%) | 0.6931 | 0.0794 | 1.0828 |

Legend – MSE: mean squared error, LP: linear predictor, CITL: calibration in the large, M-Imp: mean imputation, JMI: joint modelling imputation, JMI$^{aux}$: joint modelling imputation with auxiliary variables, SBP: systolic blood pressure, TC: total cholesterol, HDL-c: HDL-cholesterol, AD: antihypertensive drug, DM: diabetes mellitus.

**Appendix B simulation 2: imputing multiple missing predictor values scenarios using external data**

| # | Variables missing | Methods | MSE of the LP (% difference to M-Imp) | C-index | CITL | Calibration slope |
|---|---|---|---|---|---|---|
| | *Apparent performance (reference)* | | | 0.7051 | -0.0001 | 0.9999 |
| 1 | (1) SBP, (2) smoking | M-Imp | 0.0802 | 0.6908 | 0.1227 | 0.9415 |
| | | JMI | 0.0782 (-2.56%) | 0.6911 | 0.1018 | 0.9269 |
| | | JMI$^{aux}$ | 0.0801 (0.001%) | 0.6902 | 0.1123 | 0.9251 |
| 2 | (1) TC, (2) HDL-c | M-Imp | 0.0280 | 0.7005 | 0.0618 | 0.9766 |
| | | JMI | 0.0251 (-10.13%) | 0.7010 | 0.0244 | 0.9639 |
| | | JMI$^{aux}$ | 0.0248 (-11.35%) | 0.7017 | 0.0155 | 0.9724 |
| 3 | (1) TC, (2) HDL-c, (3) SBP | M-Imp | 0.0343 | 0.6994 | 0.0715 | 1.0003 |
| | | JMI | 0.0325 (-5.14%) | 0.6995 | 0.0308 | 0.9629 |
| | | JMI$^{aux}$ | 0.0324 (-5.29%) | 0.7003 | 0.0192 | 0.9703 |
| 4 | (1) TC, (2) HDL-c, (3) SBP, (4) AD | M-Imp | 0.0654 | 0.6982 | 0.1943 | 1.0217 |
| | | JMI | 0.0615 (-5.95%) | 0.6978 | 0.1661 | 0.9756 |
| | | JMI$^{aux}$ | 0.0583 (-10.91%) | 0.6998 | 0.1530 | 0.9881 |
| 5 | (1) TC, (2) HDL-c, (3) AD (4) smoking, (5) DM | M-Imp | 0.1930 | 0.6797 | 0.3090 | 0.9199 |
| | | JMI | 0.1806 (-6.42%) | 0.6803 | 0.2797 | 0.9067 |
| | | JMI$^{aux}$ | 0.1591 (-17.57%) | 0.6844 | 0.2595 | 0.9475 |
| 6 | (1) TC, (2) HDL-c, (3) AD, (4) smoking, (5) DM, (6) SBP | M-Imp | 0.1974 | 0.6785 | 0.3189 | 0.9431 |
| | | JMI | 0.1914 (-3.01%) | 0.6788 | 0.2889 | 0.9045 |
| | | JMI$^{aux}$ | 0.1664 (-15.70%) | 0.6831 | 0.2663 | 0.9483 |
| 7 | (1) Age, (2) gender, (3) TC, (4) HDL-c, (5) AD, (6) smoking, (7) DM, (8) SBP | M-Imp | 0.9167 | 0.5157 | 0.2332 | 86.984 |
| | | JMI | 0.9167 (0.00%) | 0.5157 | 0.2332 | 86.984 |
| | | JMI$^{aux}$ | 0.7269 (-20.70%) | 0.6589 | -0.0783 | 0.9687 |
| 8 | (1) Age, (2) gender | M-Imp | 0.8334 | 0.6064 | -0.1037 | 0.8230 |
| | | JMI | 0.7963 (-4.45%) | 0.6116 | -0.2221 | 0.5769 |
| | | JMI$^{aux}$ | 0.7018 (-15.79%) | 0.6721 | -0.3649 | 0.8453 |

Legend – MSE: mean squared error, LP: linear predictor, CITL: calibration in the large, M-Imp: mean imputation, JMI: joint modelling imputation, JMI$^{aux}$: joint modelling imputation with auxiliary variables, SBP: systolic blood pressure, TC: total cholesterol, HDL-c: HDL-cholesterol, AD: antihypertensive drug, DM: diabetes mellitus.

**Appendix C imputing multiple missing predictor values scenarios using enriched external data; simulation 3: scenario 1**

| # | Variables missing | Methods | MSE of the LP (% difference to M-Imp) | C-index | CITL | Calibration slope |
|---|---|---|---|---|---|---|
| | *Apparent performance* | | | 0.7051 | -0.0001 | 0.9999 |
| 1 | MD scenario 1* (i.e. without local data, reference) | M-Imp | 0.0802 | 0.6908 | 0.1227 | 0.9415 |
| | | JMI | 0.0782 (-2.56%) | 0.6911 | 0.1018 | 0.9269 |
| | | JMI+ | 0.0801 (0.001%) | 0.6902 | 0.1123 | 0.9251 |
| 2 | +100 local patients | M-Imp | 0.0794 | 0.6908 | 0.1169 | 0.9402 |
| | | JMI | 0.0769 (-3.25%) | 0.6912 | 0.0915 | 0.9252 |
| | | JMI+ | 0.0780 (-1.79%) | 0.6904 | 0.0987 | 0.9246 |
| 3 | +300 local patients | M-Imp | 0.0792 | 0.6902 | 0.1190 | 0.9393 |
| | | JMI | 0.0763 (-3.80%) | 0.6904 | 0.0910 | 0.9242 |
| | | JMI+ | 0.0771 (-2.72%) | 0.6897 | 0.0966 | 0.9229 |
| 4 | +750 local patients | M-Imp | 0.0765 | 0.6902 | 0.1091 | 0.9396 |
| | | JMI | 0.0733 (-4.37%) | 0.6904 | 0.0710 | 0.9233 |
| | | JMI+ | 0.0725 (-5.52%) | 0.6902 | 0.0693 | 0.9268 |
| 5 | +1500 local patients | M-Imp | 0.0746 | 0.6909 | 0.0845 | 0.9393 |
| | | JMI | 0.0718 (-3.90%) | 0.6913 | 0.0510 | 0.9278 |
| | | JMI+ | 0.0708 (-5.37%) | 0.6911 | 0.0485 | 0.9315 |
| 6 | +5000 local patients | M-Imp | 0.0713 | 0.6869 | 0.0779 | 0.9420 |
| | | JMI | 0.0699 (-2.00%) | 0.6874 | 0.0545 | 0.9389 |
| | | JMI+ | 0.0683 (-4.39%) | 0.6875 | 0.0482 | 0.9464 |
| 7 | +10000 local patients | M-Imp | 0.0704 | 0.6732 | 0.0863 | 0.8778 |
| | | JMI | 0.0686 (-2.65%) | 0.6733 | 0.0743 | 0.8797 |
| | | JMI+ | 0.0671 (-4.92%) | 0.6739 | 0.0681 | 0.8887 |

Legend – MSE: mean squared error, LP: linear predictor, CITL: calibration in the large, M-Imp: mean imputation, JMI: joint modelling imputation, JMIaux: joint modelling imputation with auxiliary variables, *(1) systolic blood pressure, (2) smoking.

## Appendix C (cont.) simulation 3: scenario 5

| # | Variables missing | Methods | MSE of the LP (% difference to M-Imp) | C-index | CITL | Calibration slope |
|---|---|---|---|---|---|---|
| | *Apparent performance* | | | 0.7051 | -0.0001 | 0.9999 |
| 1 | MD scenario 5* (i.e. without local data, reference) | M-Imp | 0.1974 | 0.6785 | 0.3189 | 0.9431 |
| | | JMI | 0.1914 (-3.01%) | 0.6788 | 0.2889 | 0.9045 |
| | | JMI+ | 0.1664 (-15.70%) | 0.6831 | 0.2663 | 0.9483 |
| 2 | +100 local patients | M-Imp | 0.1929 | 0.6785 | 0.3081 | 0.9418 |
| | | JMI | 0.1865 (-3.33%) | 0.6788 | 0.2775 | 0.9046 |
| | | JMI+ | 0.1605 (-16.81%) | 0.6836 | 0.2509 | 0.9502 |
| 3 | +300 local patients | M-Imp | 0.1899 | 0.6785 | 0.3054 | 0.9464 |
| | | JMI | 0.1827 (-3.78%) | 0.6788 | 0.2749 | 0.9114 |
| | | JMI+ | 0.1557 (-18.01%) | 0.6834 | 0.2445 | 0.9557 |
| 4 | +750 local patients | M-Imp | 0.1794 | 0.6787 | 0.2851 | 0.9506 |
| | | JMI | 0.1714 (-4.42%) | 0.6790 | 0.2521 | 0.9167 |
| | | JMI+ | 0.1425 (-20.56%) | 0.6844 | 0.2105 | 0.9607 |
| 5 | +1500 local patients | M-Imp | 0.1683 | 0.6790 | 0.2418 | 0.9387 |
| | | JMI | 0.1603 (-4.78%) | 0.6792 | 0.2078 | 0.9095 |
| | | JMI+ | 0.1334 (-20.72%) | 0.6851 | 0.1677 | 0.9573 |
| 6 | +5000 local patients | M-Imp | 0.1472 | 0.6736 | 0.1960 | 0.9377 |
| | | JMI | 0.1424 (-3.31%) | 0.6737 | 0.1696 | 0.9166 |
| | | JMI+ | 0.1206 (-18.11%) | 0.6789 | 0.1292 | 0.9585 |
| 7 | +10000 local patients | M-Imp | 0.1454 | 0.6630 | 0.1832 | 0.8820 |
| | | JMI | 0.1410 (-3.06%) | 0.6629 | 0.1603 | 0.8668 |
| | | JMI+ | 0.1199 (-17.54%) | 0.6703 | 0.1291 | 0.9201 |

Legend – MSE: mean squared error, LP: linear predictor, CITL: calibration in the large, M-Imp: mean imputation, JMI: joint modelling imputation, JMI$^{aux}$: joint modelling imputation with auxiliary variables, *(1) systolic blood pressure, (2) total cholesterol, (3) HDL-cholesterol, (4) smoking, (5) antihypertensive drugs, (6) Diabetes mellitus.

## Appendix C (cont.) simulation 3: scenario 8

| # | Variables missing | Methods | MSE of the LP (% difference to M-Imp) | C-index | CITL | Calibration slope |
|---|---|---|---|---|---|---|
| | *Apparent performance* | | | 0.7051 | -0.0001 | 0.9999 |
| 1 | MD scenario 8* (i.e. without local data, reference) | M-Imp | 0.8334 | 0.6064 | -0.1037 | 0.8230 |
| | | JMI | 0.7963 (-4.45%) | 0.6116 | -0.2221 | 0.5769 |
| | | JMI+ | 0.7018 (-15.79%) | 0.6721 | -0.3649 | 0.8453 |
| 2 | +100 local patients | M-Imp | 0.8281 | 0.6064 | -0.0965 | 0.8178 |
| | | JMI | 0.7787 (-6.34%) | 0.6122 | -0.1910 | 0.5893 |
| | | JMI+ | 0.6764 (-22.43%) | 0.6733 | -0.3225 | 0.8608 |
| 3 | +300 local patients | M-Imp | 0.8173 | 0.6044 | -0.0803 | 0.8118 |
| | | JMI | 0.7677 (-6.46%) | 0.6103 | -0.1662 | 0.5841 |
| | | JMI+ | 0.6510 (-25.55%) | 0.6734 | -0.2786 | 0.8665 |
| 4 | +750 local patients | M-Imp | 0.8089 | 0.6050 | -0.0540 | 0.8005 |
| | | JMI | 0.7455 (-8.50%) | 0.6108 | -0.1166 | 0.6132 |
| | | JMI+ | 0.6117 (-32.24%) | 0.6778 | -0.2138 | 0.9232 |
| 5 | +1500 local patients | M-Imp | 0.7923 | 0.6107 | -0.0205 | 0.8429 |
| | | JMI | 0.7253 (-9.24%) | 0.6131 | -0.0659 | 0.6480 |
| | | JMI+ | 0.5740 (-38.03%) | 0.6856 | -0.1451 | 0.9654 |
| 6 | +5000 local patients | M-Imp | 0.7461 | 0.6144 | 0.0905 | 0.9096 |
| | | JMI | 0.6695 (-11.44%) | 0.6202 | 0.0387 | 0.7332 |
| | | JMI+ | 0.5094 (-46.47%) | 0.6914 | -0.0503 | 1.0142 |
| 7 | +10000 local patients | M-Imp | 0.7560 | 0.6024 | 0.1577 | 0.7556 |
| | | JMI | 0.6734 (-12.27%) | 0.6065 | 0.1049 | 0.6707 |
| | | JMI+ | 0.4999 (-51.23% | 0.6973 | 0.0328 | 1.0664 |

Legend – MSE: mean squared error, LP: linear predictor, CITL: calibration in the large, M-Imp: mean imputation, JMI: joint modelling imputation, JMI$^{aux}$: joint modelling imputation with auxiliary variables, *(1) age, (2) gender.

**Appendix D: Correlation matrix (with additional patient variables) – left: local data (SMART), right: external data (UCC)**

**Appendix E – R code**

The completed UCC-CVRM data is available from *knn1*. The completed UCC-SMART data is available from *smart*. These data frames were used in the various simulation studies as follows:

```
load("knn1.RData")
load("smart.RData")
source("functions.r")

# test imputation model
frh_vars <-
c("leeftijd","geslacht","labchol","labhdl","bdsys","mht_all","roken","vz_
DM")
testMSE(ds[,frh_vars],missing_var="labchol",n.imp=1,method="internal",see
d=1221)

apparent_performance(data=data_imp) # c = 0.7061156, intercept = -
0.0000517122, slope = 0.9999569, eo = 1.000052

# optimism corrected performance
# age_gender_perf <- oc_performance(data=smart)
oc_performance(data=smart)

apparent_performance(data=smart)
oc_performance(data=smart)

##################################################
# 1. Simulation study without ldl cholesterol #
##################################################

smart.chol <- smart
smart.chol[,"ldlchol"] <- NULL

ucc.chol <- ucc
ucc.chol[,"ldlchol"] <- NULL

scenarios <- list(scen1 = c("labchol"),
         scen2 = c("labhdl"),
         scen3 = c("labchol","labhdl"),
         scen4 = c("labchol","labhdl","roken","mht_all","vz_DM"),
         scen5 = c("labchol","labhdl","bdsys","mht_all","roken","vz_DM"),
         scen6 = c("labchol","labhdl","bdsys"),
         scen7 = c("labchol","labhdl","bdsys","mht_all"),
         scen8 =
c("leeftijd","geslacht","labchol","labhdl","bdsys","mht_all","roken","vz_
DM"))

sim_1_smart <- run_simulation(ref_data  = smart.chol,
              scenarios = scenarios,
              validation = "jackknife",
              seed     = 12345)

sim_1_ucc  <- run_simulation(ref_data  = smart.chol,
              ext_data  = ucc.chol,
              scenarios = scenarios,
              validation = "external",
              seed     = 12345)
```

```
#############################
# Results simulation study #
#############################

eval_jk(results=sim_1_smart,imputation_method=1,pattern=1)
eval_jk(results=sim_1_smart,imputation_method=2,pattern=1)
eval_jk(results=sim_1_smart,imputation_method=3,pattern=1)
eval_jk(results=sim_1_ucc,imputation_method=1,pattern=1)
eval_jk(results=sim_1_ucc,imputation_method=2,pattern=1)
eval_jk(results=sim_1_ucc,imputation_method=3,pattern=1)

eval_jk(results=sim_1_smart,imputation_method=1,pattern=2)
eval_jk(results=sim_1_smart,imputation_method=2,pattern=2)
eval_jk(results=sim_1_smart,imputation_method=3,pattern=2)
eval_jk(results=sim_1_ucc,imputation_method=1,pattern=2)
eval_jk(results=sim_1_ucc,imputation_method=2,pattern=2)
eval_jk(results=sim_1_ucc,imputation_method=3,pattern=2)

eval_jk(results=sim_1_smart,imputation_method=1,pattern=3)
eval_jk(results=sim_1_smart,imputation_method=2,pattern=3)
eval_jk(results=sim_1_smart,imputation_method=3,pattern=3)
eval_jk(results=sim_1_ucc,imputation_method=1,pattern=3)
eval_jk(results=sim_1_ucc,imputation_method=2,pattern=3)
eval_jk(results=sim_1_ucc,imputation_method=3,pattern=3)

eval_jk(results=sim_1_smart,imputation_method=1,pattern=4)
eval_jk(results=sim_1_smart,imputation_method=2,pattern=4)
eval_jk(results=sim_1_smart,imputation_method=3,pattern=4)
eval_jk(results=sim_1_ucc,imputation_method=1,pattern=4)
eval_jk(results=sim_1_ucc,imputation_method=2,pattern=4)
eval_jk(results=sim_1_ucc,imputation_method=3,pattern=4)

eval_jk(results=sim_1_smart,imputation_method=1,pattern=5)
eval_jk(results=sim_1_smart,imputation_method=2,pattern=5)
eval_jk(results=sim_1_smart,imputation_method=3,pattern=5)
eval_jk(results=sim_1_ucc,imputation_method=1,pattern=5)
eval_jk(results=sim_1_ucc,imputation_method=2,pattern=5)
eval_jk(results=sim_1_ucc,imputation_method=3,pattern=5)

eval_jk(results=sim_1_smart,imputation_method=1,pattern=6)
eval_jk(results=sim_1_smart,imputation_method=2,pattern=6)
eval_jk(results=sim_1_smart,imputation_method=3,pattern=6)
eval_jk(results=sim_1_ucc,imputation_method=1,pattern=6)
eval_jk(results=sim_1_ucc,imputation_method=2,pattern=6)
eval_jk(results=sim_1_ucc,imputation_method=3,pattern=6)

eval_jk(results=sim_1_smart,imputation_method=1,pattern=7)
eval_jk(results=sim_1_smart,imputation_method=2,pattern=7)
eval_jk(results=sim_1_smart,imputation_method=3,pattern=7)
eval_jk(results=sim_1_ucc,imputation_method=1,pattern=7)
eval_jk(results=sim_1_ucc,imputation_method=2,pattern=7)
eval_jk(results=sim_1_ucc,imputation_method=3,pattern=7)

eval_jk(results=sim_1_smart,imputation_method=1,pattern=8)
eval_jk(results=sim_1_smart,imputation_method=2,pattern=8)
eval_jk(results=sim_1_smart,imputation_method=3,pattern=8)
eval_jk(results=sim_1_ucc,imputation_method=1,pattern=8)
eval_jk(results=sim_1_ucc,imputation_method=2,pattern=8)
eval_jk(results=sim_1_ucc,imputation_method=3,pattern=8)
```

```
##################################################
# 2. Simulation study with remaining scenarios #
##################################################

scenarios <- list(scen1 = c("bdsys"),
                  scen2 = c("mht_all"),
                  scen3 = c("roken"),
                  scen4 = c("vz_DM"),
                  scen5 = c("roken","mht_all"),
                  scen6 = c("bdsys","mht_all","roken"),
                  scen7 = c("leeftijd","geslacht"))

sim_2_smart <- run_simulation(ref_data  = smart,
                 scenarios = scenarios,
                 validation = "jackknife",
                 seed      = 12345)
sim_2_ucc   <- run_simulation(ref_data  = smart,
                 ext_data  = ucc,
                 scenarios = scenarios,
                 validation = "external",
                 seed      = 12345)

#############################
# Results simulation study #
#############################

eval_jk(results=sim_2_smart,imputation_method=1,pattern=1)
eval_jk(results=sim_2_smart,imputation_method=2,pattern=1)
eval_jk(results=sim_2_smart,imputation_method=3,pattern=1)
eval_jk(results=sim_2_ucc,imputation_method=1,pattern=1)
eval_jk(results=sim_2_ucc,imputation_method=2,pattern=1)
eval_jk(results=sim_2_ucc,imputation_method=3,pattern=1)

eval_jk(results=sim_2_smart,imputation_method=1,pattern=2)
eval_jk(results=sim_2_smart,imputation_method=2,pattern=2)
eval_jk(results=sim_2_smart,imputation_method=3,pattern=2)
eval_jk(results=sim_2_ucc,imputation_method=1,pattern=2)
eval_jk(results=sim_2_ucc,imputation_method=2,pattern=2)
eval_jk(results=sim_2_ucc,imputation_method=3,pattern=2)

eval_jk(results=sim_2_smart,imputation_method=1,pattern=3)
eval_jk(results=sim_2_smart,imputation_method=2,pattern=3)
eval_jk(results=sim_2_smart,imputation_method=3,pattern=3)
eval_jk(results=sim_2_ucc,imputation_method=1,pattern=3)
eval_jk(results=sim_2_ucc,imputation_method=2,pattern=3)
eval_jk(results=sim_2_ucc,imputation_method=3,pattern=3)

eval_jk(results=sim_2_smart,imputation_method=1,pattern=4)
eval_jk(results=sim_2_smart,imputation_method=2,pattern=4)
eval_jk(results=sim_2_smart,imputation_method=3,pattern=4)
eval_jk(results=sim_2_ucc,imputation_method=1,pattern=4)
eval_jk(results=sim_2_ucc,imputation_method=2,pattern=4)
eval_jk(results=sim_2_ucc,imputation_method=3,pattern=4)

eval_jk(results=sim_2_smart,imputation_method=1,pattern=5)
eval_jk(results=sim_2_smart,imputation_method=2,pattern=5)
eval_jk(results=sim_2_smart,imputation_method=3,pattern=5)
eval_jk(results=sim_2_ucc,imputation_method=1,pattern=5)
```

```r
eval_jk(results=sim_2_ucc,imputation_method=2,pattern=5)
eval_jk(results=sim_2_ucc,imputation_method=3,pattern=5)

eval_jk(results=sim_2_smart,imputation_method=1,pattern=6)
eval_jk(results=sim_2_smart,imputation_method=2,pattern=6)
eval_jk(results=sim_2_smart,imputation_method=3,pattern=6)
eval_jk(results=sim_2_ucc,imputation_method=1,pattern=6)
eval_jk(results=sim_2_ucc,imputation_method=2,pattern=6)
eval_jk(results=sim_2_ucc,imputation_method=3,pattern=6)

eval_jk(results=sim_2_smart,imputation_method=1,pattern=7)
eval_jk(results=sim_2_smart,imputation_method=2,pattern=7)
eval_jk(results=sim_2_smart,imputation_method=3,pattern=7)
eval_jk(results=sim_2_ucc,imputation_method=1,pattern=7)
eval_jk(results=sim_2_ucc,imputation_method=2,pattern=7)
eval_jk(results=sim_2_ucc,imputation_method=3,pattern=7)

 ################################################
# 3. simulation study with enriched UCC data #
################################################

# when chol is part of the scenario:
smart.chol <- smart
ucc.chol <- ucc
smart.chol[,"ldlchol"] <- NULL
ucc.chol[,"ldlchol"] <- NULL

# separate common variables in smart & ucc
smart_comvar <- smart.chol[,which(names(smart.chol) %in%
names(ucc.chol))]
ucc_comvar <- ucc.chol[,which(names(ucc.chol) %in% names(smart.chol))]

# set aside different lenghts 'local' data from smart
set.seed(12345)
add1 <- smart_comvar[sample(nrow(smart_comvar),100),]
add2 <- smart_comvar[sample(nrow(smart_comvar),300),]
add3 <- smart_comvar[sample(nrow(smart_comvar),750),]
add4 <- smart_comvar[sample(nrow(smart_comvar),1500),]
add5 <- smart_comvar[sample(nrow(smart_comvar),5000),]
add6 <- smart_comvar[sample(nrow(smart_comvar),10000),]

# enrich ucc data with different lenghts 'local' data from smart
ucc1 <- rbind(ucc_comvar,add1)
ucc2 <- rbind(ucc_comvar,add2)
ucc3 <- rbind(ucc_comvar,add3)
ucc4 <- rbind(ucc_comvar,add4)
ucc5 <- rbind(ucc_comvar,add5)
ucc6 <- rbind(ucc_comvar,add6)

# remove data used for enrichment from smart
smart1 <- smart_comvar[-as.numeric(rownames(add1)),]
smart2 <- smart_comvar[-as.numeric(rownames(add2)),]
smart3 <- smart_comvar[-as.numeric(rownames(add3)),]
smart4 <- smart_comvar[-as.numeric(rownames(add4)),]
smart5 <- smart_comvar[-as.numeric(rownames(add5)),]
smart6 <- smart_comvar[-as.numeric(rownames(add6)),]

# add back time/status variables
smart1$time <- smart[as.numeric(rownames(smart1)),"time"]
smart1$status <- smart[as.numeric(rownames(smart1)),"status"]
```

```r
smart2$time   <- smart[as.numeric(rownames(smart2)),"time"]
smart2$status <- smart[as.numeric(rownames(smart2)),"status"]
smart3$time   <- smart[as.numeric(rownames(smart3)),"time"]
smart3$status <- smart[as.numeric(rownames(smart3)),"status"]
smart4$time   <- smart[as.numeric(rownames(smart4)),"time"]
smart4$status <- smart[as.numeric(rownames(smart4)),"status"]
smart5$time   <- smart[as.numeric(rownames(smart5)),"time"]
smart5$status <- smart[as.numeric(rownames(smart5)),"status"]
smart6$time   <- smart[as.numeric(rownames(smart6)),"time"]
smart6$status <- smart[as.numeric(rownames(smart6)),"status"]

rm(add1,add2,add3,add4,add5,add6,smart_comvar,ucc_comvar,smart.chol,ucc.chol)

# scenarios with chol
scenarios <- list(scen1 =
c("bdsys","labchol","labhdl","roken","mht_all","vz_DM"))

# scenarios without chol
scenarios <- list(scen1 = c("bdsys","roken"))

sim_3_enri1  <- run_simulation(ref_data = smart1,
                ext_data  = ucc1,
                model_data = smart,
                scenarios = scenarios,
                validation = "enrich",
                seed    = 12345)
sim_3_enri2  <- run_simulation(ref_data = smart2,
                ext_data  = ucc2,
                model_data = smart,
                scenarios = scenarios,
                validation = "enrich",
                seed    = 12345)
sim_3_enri3  <- run_simulation(ref_data = smart3,
                ext_data  = ucc3,
                model_data = smart,
                scenarios = scenarios,
                validation = "enrich",
                seed    = 12345)
sim_3_enri4  <- run_simulation(ref_data = smart4,
                ext_data  = ucc4,
                model_data = smart,
                scenarios = scenarios,
                validation = "enrich",
                seed    = 12345)
sim_3_enri5  <- run_simulation(ref_data = smart5,
                ext_data  = ucc5,
                model_data = smart,
                scenarios = scenarios,
                validation = "enrich",
                seed    = 12345)
sim_3_enri6  <- run_simulation(ref_data = smart6,
                ext_data  = ucc6,
                model_data = smart,
                scenarios = scenarios,
                validation = "enrich",
                seed    = 12345)
save(sim_3_enri4, file="simulation_3_age_gender.RData")

###########################
```

```
# Results simulation study #
#############################

eval_jk(results=sim_3_enri1,imputation_method=1,pattern=1)
eval_jk(results=sim_3_enri1,imputation_method=2,pattern=1)
eval_jk(results=sim_3_enri1,imputation_method=3,pattern=1)

eval_jk(results=sim_3_enri2,imputation_method=1,pattern=1)
eval_jk(results=sim_3_enri2,imputation_method=2,pattern=1)
eval_jk(results=sim_3_enri2,imputation_method=3,pattern=1)

eval_jk(results=sim_3_enri3,imputation_method=1,pattern=1)
eval_jk(results=sim_3_enri3,imputation_method=2,pattern=1)
eval_jk(results=sim_3_enri3,imputation_method=3,pattern=1)

eval_jk(results=sim_3_enri4,imputation_method=1,pattern=1)
eval_jk(results=sim_3_enri4,imputation_method=2,pattern=1)
eval_jk(results=sim_3_enri4,imputation_method=3,pattern=1)

eval_jk(results=sim_3_enri5,imputation_method=1,pattern=1)
eval_jk(results=sim_3_enri5,imputation_method=2,pattern=1)
eval_jk(results=sim_3_enri5,imputation_method=3,pattern=1)

eval_jk(results=sim_3_enri6,imputation_method=1,pattern=1)
eval_jk(results=sim_3_enri6,imputation_method=2,pattern=1)
eval_jk(results=sim_3_enri6,imputation_method=3,pattern=1)
```

The content of the file *functions.r* is as follows:

```
###########
# Authors #
###########

# Steven Nijman
# Jeroen Hoogland
# Thomas Debray

########################
# Package Requirements #
########################

library(foreign)
library(mice)     # version 3.6.0
library(condMVNorm) # version 2015.2-1
library(survival)

########################

#' Single patient joint imputation
#'
#' @author Steven W J Nijman \email{S.W.J.Nijman@@umcutrecht.nl}
#' @author Jeroen Hoogland  \email{J.Hoogland-2@@umcutrecht.nl}
#' @author Thomas P A Debray \email{T.Debray@@umcutrecht.nl}
#'
#' @param data single patient data
#' @param imp_means prior estimated column means of data
#' @param imp_cov prior estimated covariance matrix of data
#' @param n.imp number of imputations to make; default==1, which means it
uses expected values
```

```r
#'
#' @return Returns single patient data with imputed values
impJoint <-
function(data=data,imp_means=imp_means,imp_cov=imp_cov,n.imp=1,...) {
 if (class(data) != "data.frame") {
  stop ("Data object should be a data frame")
 }
 if (nrow(data) > 1) {
  stop ("Data should contain a single patient")
 }
 if ("time" %in% colnames(data)) {
  stop ("No time info should be used for imputation")
 }
 if ("status" %in% colnames(data)) {
  stop ("No event info should be used for imputation")
 }

 dep  <- which(is.na(data))
 given <- which(!is.na(data))
 depnames <- colnames(data[dep])
 givennames <- colnames(data[given])
 missing.col <- which(is.na(data))
 data <- as.matrix(data)

 if(length(given)==0) {
  data_imp <- t(as.data.frame(imp_means))
 } else if(length(given)>0) {
  # Extract conditional Mean and conditional var
  cond <-
condMVN(mean=imp_means,sigma=imp_cov,dep=dep,given=given,X=data[given])

  if (n.imp == 1) {
   # Just impute the expected value if we only need 1 imputation
   x.imp <- matrix(cond$condMean, nrow = 1)
  } else if (n.imp < 1000) {
   # We should not use multiple imputation of the number of imputed
values is very low.
   # The empirical covariance of imputed values is very unreliable in
such circumstances
   stop ("A minimum of 1000 imputations should be generated when drawing
random samples, instead of using the conditional mean.")
  } else {
   # Draw from a multivariate normal if multiple imputations required
   x.imp <- mvrnorm(n = n.imp, mu = cond$condMean, Sigma = cond$condVar,
tol = 1e-6, empirical = FALSE, EISPACK = FALSE)
  }

  x.obs <- matrix(data[given], nrow=n.imp, ncol=length(given), byrow =
TRUE)
  data_imp <- as.data.frame(cbind(x.obs, x.imp))
  colnames(data_imp) <- c(givennames, depnames)
 }
 data.frame(data_imp)
}

# calculate mse of imputation methods
testMSE <-
function(data=data,missing_var=missing_var,n.imp=n.imp,method="jackknife"
,seed=12345,...) {
 jmimpdat  <- meanimpdat <- data
```

```r
  jmimpdat[,] <- meanimpdat[,] <- NA
 if (!is.na(seed)) {
  set.seed(seed)
 }
 pb <- txtProgressBar(min = 0, max = nrow(data), style = 3)
 for (i in 1:nrow(data)) {
  setTxtProgressBar(pb, i)

  if (method == "jackknife") {
   training_data <- data[-i,]
   test_case    <- data[i,]
  } else if (method == "internal") {
   training_data <- data
   test_case    <- data[i,]
  } else if (method == "external") {
   training_data <- model_data
   test_case    <- data[i,]
  } else {
   stop ("Validation method not supported")
  }

  mu   <- colMeans(training_data)
  sigma <- cov(training_data)

  test_case[,missing_var] <- NA

  meanimpdat[i,] <- jmimpdat[i,] <- test_case
  meanimpdat[i,missing_var] <- mu[missing_var]
  jmimpdat[i,missing_var]   <-
impJoint(data=test_case,imp_means=mu,imp_cov=sigma)[missing_var]
 }
 close(pb)
 if(length(missing_var)==1) {
  result <- data.frame(mse_meanimp = mean((meanimpdat[,missing_var] -
data[,missing_var])**2),
            mse_jmimp  = mean((jmimpdat[,missing_var] -
data[,missing_var])**2))
 } else if(length(missing_var>=2)) {
  result <- data.frame(mse_meanimp = colMeans((meanimpdat[,missing_var] -
data[,missing_var])**2),
            mse_jmimp  = colMeans((jmimpdat[,missing_var] -
data[,missing_var])**2))
 }

 return(result)
}

# calibration intercept
cal_intercept <- function(model, data) {
 p <- log(predict(model, newdata = data, type="expected")) # Expected
number of events
 fit1 <- glm(status ~ offset(p), family="poisson", data = data)
 coef(fit1) # Should be 0
}

# calibration slope
cal_slope <- function(model, data) {
 p <- log(predict(model, newdata = data, type="expected"))
```

```r
 # Calculate the linear predictor
 # lp <- model.matrix(model$formula,data)[,-1] %*% model$coefficients
 lp <- matrix(data$lp) %*% model$coefficients
 lpc <- predict(model, newdata = data, type="lp") #Centered linear
predictor
 logbase <- p - lp
 fit2 <- glm(status ~ lpc + offset(logbase), family = poisson, data =
data)
 coef(fit2)["lpc"] # Should be 1
}

# ten year risk
ten_risk <- function(model=model,lp=lp) {
 base <- basehaz(fit,centered=F)
 timediff <- (abs(base$time-365*10)) # Identify patient with 10y follow-
up
 base10 <- base$hazard[which(timediff == min(timediff))] # Extract
cumulative baseline hazard for a patient with 10y follow-up
 predsurv <- exp(-base10)**(exp(lp))
 predrisk <- 1-predsurv

 # risk <- rep(NA,nrow(ds))
 # bh   <- summary(survfit(Surv(time,status)~1,data=ds),time=3650)$surv #
10yr cumulative survival
 # for(i in 1:nrow(ds)) risk[i] <- 1-bh^exp(lp[i])

 return(predrisk)
}

# apparent performance of prediction model
apparent_performance <- function(data=data,...) {
 # fit <-
coxph(Surv(time,status)~leeftijd+geslacht+labchol+labhdl+bdsys+mht_all+ro
ken+vz_DM, data=data)
 fit <- coxph(Surv(time,status)~leeftijd+geslacht, data=data)

 # Calculate the linear predictor
 lp  <- model.matrix(fit$formula,data)[,-1] %*% fit$coefficients

 # c-index
 cindex <-
as.numeric(survConcordance(Surv(time,status)~lp,data=data)$concordance)

 # calibration intercept
 intercept <- cal_intercept(fit,data)

 # calibration slope
 slope   <- cal_slope(fit,data)

 # EO
 eo  <- sum(predict(fit,newdata=data,type="expected"))/sum(data$status)

 return(c(cindex=cindex,
      cal_inter=intercept,
      cal_slope=slope,
      eo=eo))
}

# optimism corrected performance of prediction model
```

```r
oc_performance <- function(data=data,...) {
 result <- array(NA,dim=c(nrow(data),3))
 colnames(result) <- c("lp","time","status")
 result[,c("time","status")] <- c(data[,"time"],data[,"status"])

 pb <- txtProgressBar(min = 0, max = nrow(data), style = 3)
 for(i in 1:nrow(data)) {
  setTxtProgressBar(pb, i)
  training_data <- data[-i,]
  test_case <- data[i,]

  # fit <-
coxph(Surv(time,status)~leeftijd+geslacht+labchol+labhdl+bdsys+mht_all+ro
ken+vz_DM,data=training_data)
  fit <- coxph(Surv(time,status)~leeftijd+geslacht,data=training_data)

  # Calculate the linear predictor
  result[i,"lp"]  <- model.matrix(fit$formula,test_case)[,-1] %*%
fit$coefficients
 }
 result <- as.data.frame(result)

 # refit cox
 fit2 <- coxph(Surv(time,status)~lp,data=result)

 # c-index
 cindex <-
as.numeric(survConcordance(Surv(time,status)~lp,data=result)$concordance)

 # calibration intercept
 intercept <- cal_intercept(fit2,result)

 # calibration slope
 slope   <- cal_slope(fit2,result)

 # EO
 eo  <-
sum(predict(fit2,newdata=result,type="expected"))/sum(result$status)

 return(c(cindex=cindex,
      cal_inter=intercept,
      cal_slope=slope,
      eo=eo))
}

#' Simulate missing data and impute using joint modelling imputation
#'
#' @author Steven W J Nijman \email{S.W.J.Nijman@@umcutrecht.nl}
#' @author Thomas P A Debray \email{T.Debray@@umcutrecht.nl}
#'
#' @param ref_data reference data, used to estimate prediction model in
#' @param ext_data used when validation==external, to specify data on
which imputation model should be estimated.
#' @param model_data used when validation==enrich, to specify prediction
model in jackknife sample of reference data, of which part is used to
enrich external data.
#' @param scenarios list of missing value scenarios using matching
colnames.
```

```r
#' @param validation specifies which method is used for validation,
defalut is "jackknife". Options are internal, external, jackknife or
enrich.
#' @param seed set seed for stochastic processes in simulation.
#'
#' @return Returns a matrix with reference linear predictor, imputed
linear predictor, expected number of events, time-to-event, status,
method used, scenario imputed and row reference.
run_simulation <- function(ref_data=ref_data,
              ext_data=ext_data,
              model_data=model_data,
              scenarios=scenarios,
              validation="jackknife",
              seed=seed,...) {

 # Create large matrix based on amount of scenarios, patients in
reference data and amount of imputation methods (3)
 sim_frame <- matrix(NA,nrow=(nrow(ref_data)*length(scenarios)*3),ncol=8)
 colnames(sim_frame) <-
c("lp_ref","lp_est","Expected","time","status","scenario","method","rowref")
 sim_frame[,"scenario"] <-
rep(1:length(scenarios),each=nrow(ref_data),times=3)
 sim_frame[,"method"]   <- rep(1:3,each=nrow(ref_data)*length(scenarios))
 sim_frame[,"time"]     <- rep(ref_data$time,times=3*length(scenarios))
 sim_frame[,"status"]   <- rep(ref_data$status,times=3*length(scenarios))

 # set row reference in large matrix for correct row in reference data
 sim_frame[,"rowref"]   <- rep(1:nrow(ref_data),times=3*length(scenarios))

 # added to specify which row to take when enriching external data
 if(validation=="enrich") {
  modelref <- matrix(NA,nrow=nrow(sim_frame),ncol=1)
  colnames(modelref) <- "modelref"
  for(row in 1:nrow(sim_frame)) {
   j <- as.numeric(sim_frame[row,"rowref"])
    modelref[row] <- as.numeric(rownames(model_data[which(model_data$time
== ref_data[j,"time"] &
                                  model_data$status == ref_data[j,"status"]
&
                                  model_data$leeftijd ==
ref_data[j,"leeftijd"] &
                                  model_data$geslacht ==
ref_data[j,"geslacht"] &
                                  model_data$labchol ==
ref_data[j,"labchol"] &
                                  model_data$bdsys ==
ref_data[j,"bdsys"]),]))
   }
  sim_frame <- cbind(sim_frame, modelref)
 }

 # separate framingham predictors and all covariates (including
predictors and auxiliary variables)
 frh_vars <-
c("leeftijd","geslacht","labchol","labhdl","bdsys","mht_all","roken","vz_DM")

 # determine additional patient variables when using local or external
data
```

```r
  if(validation=="external") {
   aux_vars <- names(ref_data[which(names(ref_data) %in%
names(ext_data))])
  } else {
   aux_vars <- names(ref_data[-which(names(ref_data) %in%
c("time","status"))])
  }

 if (!is.na(seed)) {
  set.seed(seed)
 }

 pb <- txtProgressBar(min = 0, max = nrow(sim_frame), style = 3)

 # run simulation
 for (i in 1:nrow(sim_frame)) {
  setTxtProgressBar(pb, i)
  j <- sim_frame[i,"rowref"]
  if("modelref" %in% colnames(sim_frame)) k <- sim_frame[i,"modelref"]

  # validation selection
  if (validation == "jackknife") {
   training_data <- imp_data <- ref_data[-j,]
   test_case   <- ref_data[j,]
  } else if (validation == "internal") {
   training_data <- imp_data <- ref_data
   test_case   <- ref_data[j,]
  } else if (validation == "external") {
   training_data <- ref_data[-j,]
   imp_data    <- ext_data
   test_case   <- ref_data[j,]
  } else if (validation == "enrich") {
   training_data <- model_data[-k,]
   imp_data    <- ext_data
   test_case   <- ref_data[j,]
  } else {
   stop ("Validation method not supported")
  }

  # estimate cox
  fit <-
coxph(Surv(time,status)~leeftijd+geslacht+labchol+labhdl+bdsys+mht_all+ro
ken+vz_DM, data=training_data)

  # calculate lp given complete data
  sim_frame[i,"lp_ref"] <- model.matrix(fit$formula,test_case)[,-1] %*%
fit$coefficients

  # given scenario make specified predictors missing
  missing_var <- scenarios[sim_frame[i,"scenario"]][[1]]
  test_case[,missing_var] <- NA

  # estimate means and covariance based on pre-specified training data
  mu   <- colMeans(imp_data[,aux_vars])
  sigma <- cov(imp_data[,aux_vars])

  # given method specify imputation (mean, joint or joint with auxiliary)
  method_num <- as.numeric(sim_frame[i,"method"])
  if(method_num==1) {
   test_case[,missing_var]  <- mu[missing_var]
```

```r
  } else if(method_num==2) {
   test_case[,missing_var]  <- impJoint(data=test_case[,frh_vars],
                        imp_means=mu[frh_vars],
                        imp_cov=sigma[frh_vars,frh_vars])[missing_var]
  } else if(method_num==3) {
   test_case[,missing_var]  <- impJoint(data=test_case[,aux_vars],
                        imp_means=mu,
                        imp_cov=sigma)[missing_var]
  }
  # extract expected number of events
  sim_frame[i,"Expected"]  <- predict(fit, newdata = test_case, type="expected")

  # calculate estimated lp given imputed data
  sim_frame[i,"lp_est"]   <- model.matrix(fit$formula,test_case)[,-1] %*% fit$coefficients
 }
 return(as.data.frame(sim_frame))
}

eval_jk <- function(results = results,
          imputation_method = 1,
          pattern = 1,...) {
 data  <- subset(results, method==imputation_method & scenario==pattern)
 mse   <- (sum((data$lp_ref-data$lp_est)**2))/nrow(data)
 cindex <- as.numeric(survConcordance(Surv(time, status)~lp_est,
data=data)$concordance)

 # calculate required estimates (Crowson 2016)
 p    <- log(data$Expected)
 data$lpc <- data$lp_est - mean(data$lp_est)
 logbase <- p - data$lp_est

 # extract quantities of interest
 slope  <- as.numeric(glm(status ~ lpc + offset(logbase), family =
poisson, data = data)$coefficients["lpc"])
 inter  <- as.numeric(glm(status ~ offset(p), family="poisson", data =
data)$coefficients["(Intercept)"])

 # calculate base OE
 OE1 <- sum(data$Expected) / sum(data$status)
 data$group <- cut(data$lpc,quantile(data$lpc,probs=seq(0,1,0.33)))
 OE <- rep(NA,length(unique(data$group)))
 for(i in 1:length(unique(data$group))) {
  OE[i] <- sum(data$Expected[which(data$group==levels(data$group)[i])]) /
sum(data$status[which(data$group==levels(data$group)[i])])
 }

 return(list(MSE      = mse,
       Cindex    = cindex,
       Cal_intercept = inter,
       Cal_slope   = slope,
       baseOE    = OE1,
       seqOE     = OE))
}

get_groupkm <- function(sim=sim,m=3,scen=scen) {
 km    <- list()
 for(i in 1:m) {
```

```
  data        <- subset(sim, method==i & scenario==scen)
  timediff    <- abs(data$time-365*10)
  tenyrpatient <- data[which(timediff == min(timediff)),]
  base.risk.10y <- tenyrpatient$Expected/exp(tenyrpatient$lp_ref)
  surv.10y    <- exp(-base.risk.10y)
  predrisk    <- 1 - (surv.10y^exp(data$lp_est))
  km[[i]]     <-
groupkm(predrisk,Surv(data$time,data$status),g=5,u=(10*365),pl=F)
 }
 km1  <- data.frame(km[[1]])
 km2  <- data.frame(km[[2]])
 km3  <- data.frame(km[[3]])
 data <- rbind(km1,km2,km3)
 data$method <- as.factor(c(rep("MI",5),rep("JMI",5),rep("JMI.adj",5)))
 return(data)
}
```